\documentclass[superscriptaddress,aps,showpacs,nofootinbib,twocolumn]{revtex4}

\usepackage{graphicx}
\usepackage{amsmath,amssymb}
\def\mt{\widetilde{m}_1}

\begin{document}

\title{Aspects of thermal leptogenesis in braneworld cosmology}
\author{M. C. Bento}
\email{bento@sirius.ist.utl.pt} \affiliation{Departamento de F\'{\i}sica and Centro
de F\'{\i}sica Te\'orica de Part\'{\i}culas, Instituto Superior T\'{e}cnico, Av. Rovisco
Pais, 1049-001 Lisboa, Portugal}
\author{R. Gonz\'{a}lez Felipe}
\email{gonzalez@cftp.ist.utl.pt} \affiliation{Departamento de F\'{\i}sica and Centro
de F\'{\i}sica Te\'orica de Part\'{\i}culas, Instituto Superior T\'{e}cnico, Av. Rovisco
Pais, 1049-001 Lisboa, Portugal}
\author{N. M. C. Santos}
\email{ncsantos@cftp.ist.utl.pt} \affiliation{Departamento de F\'{\i}sica and Centro
de F\'{\i}sica Te\'orica de Part\'{\i}culas, Instituto Superior T\'{e}cnico, Av. Rovisco
Pais, 1049-001 Lisboa, Portugal}

\begin{abstract}
The mechanism of thermal leptogenesis is investigated in the high-energy regime
of braneworld cosmology. Within the simplest seesaw framework with hierarchical
heavy Majorana neutrinos, we study the implications of the modified Friedmann
equation on the realization of this mechanism. In contrast with the usual
leptogenesis scenario of standard cosmology, where low-energy neutrino data
favors a mildly strong washout regime, we find that leptogenesis in the
braneworld regime is successfully realized in a weak washout regime.
Furthermore, a quasi-degenerate light neutrino mass spectrum is found to be
compatible with this scenario. For an initially vanishing heavy Majorana
neutrino abundance, thermal leptogenesis in the brane requires the decaying
heavy Majorana neutrino mass to be $M_1 \gtrsim 10^{10}$~GeV and the
fundamental five-dimensional gravity scale $10^{12} \lesssim M_5 \lesssim
10^{16}$~GeV, which corresponds to a transition from brane to standard
cosmology at temperatures $10^{8} \lesssim T_t \lesssim 10^{14}$~GeV.
\end{abstract}

%\date{\today}
\maketitle

\section{Introduction}

The most recent Wilkinson Microwave Anisotropy Probe (WMAP) results and big
bang nucleosynthesis analysis of the primordial deuterium abundance
imply~\cite{Spergel:2003cb}
\begin{equation}
\eta_{B}=\frac{n_B-n_{\bar{B}}}{n_\gamma}=(6.1\pm0.3)\times10^{-10}\,,\label{BAU}
\end{equation}
for the baryon-to-photon ratio of number densities. Among the viable mechanisms
to explain such a matter-antimatter asymmetry in the universe,
leptogenesis~\cite{Fukugita:1986hr} has become one of the most compelling
scenarios due to its simplicity and close connection with low-energy neutrino
physics. In the simplest framework, consisting on the addition of right-handed
neutrinos to the standard model, leptogenesis can be easily realized by means
of the out-of-equilibrium decays of the heavy Majorana neutrinos $N_i$ at
temperatures below their mass scale $M_i$. The lepton asymmetry generated in
the presence of $CP$-violating processes is then partially converted into a
baryon asymmetry by the sphalerons~\cite{Kuzmin:1985mm}.

Right-handed neutrinos can also provide a natural explanation for the smallness
of the neutrino masses through the well-known seesaw
mechanism~\cite{Minkowski:1977sc}. In this respect, the heavy Majorana neutrino
mass spectrum turns out to be crucial for a successful implementation of both
leptogenesis and the seesaw mechanism. In particular, the standard thermal
leptogenesis scenario with hierarchical heavy Majorana neutrino masses ($M_1
\ll M_2 < M_3$) constrains the lightest heavy Majorana mass to be $M_1 \gtrsim
4 \times 10^8$~GeV~\cite{Davidson:2002qv,Buchmuller:2002rq,Hambye:2003rt}, if
$N_1$ is in thermal equilibrium before it decays, or $M_1 \gtrsim 2 \times
10^9$~GeV~\cite{Giudice:2003jh} for a zero initial $N_{1}$ population.
Moreover, an upper bound on the light neutrino masses, $m_i <
0.12$~eV~\cite{Buchmuller:2004nz}, is implied by successful leptogenesis.

In determining the departure from thermal equilibrium, the interplay between
the expansion rate of the universe $H$ and the particle reaction rates involved
at the relevant epoch is crucial. Of course, such an interplay depends on the
specific properties of early cosmology. Braneworld cosmology (BC) has opened up
the possibility for a new phenomenology of the early universe. In particular,
the Randall-Sundrum type II braneworld model~\cite{Randall:1999vf} has recently
received much attention. In this model, the expansion dynamics at early epochs
is changed by the presence in the Friedmann equation of a term quadratic in the
energy density~\cite{Binetruy:1999ut},
\begin{align}
\label{Fried} H^2 = \frac{8\pi}{3 M_P^2} ~ \rho ~ \left(1 +
\frac{\rho}{2\lambda} \right) ~,\quad \lambda  = \frac{3}{4 \pi}
\frac{M_5^6}{M_P^2}~,
\end{align}
where $M_P \simeq 1.22 \times 10^{19}$~GeV is the four-dimensional Planck mass,
$M_5$ is the five-dimensional Planck mass and we have set the four-dimensional
cosmological constant to zero and assumed that inflation rapidly makes any dark
radiation term negligible. Notice that Eq.~(\ref{Fried}) reduces to the usual
Friedmann equation at sufficiently low energies, \mbox{$\rho \ll \lambda$},
while at very high energies we have $H\propto\rho$. This behavior has important
consequences on early universe phenomena such as
inflation~\cite{Maartens:1999hf} and the generation of the baryon
asymmetry~\cite{Mazumdar:2000gj}.

In this paper, we study how the standard mechanism of thermal leptogenesis is
affected by modifications in the expansion rate of the universe due to early
braneworld cosmology. We perform a thorough analysis of this scenario by
solving the relevant Boltzmann equations following the approach of
Refs.~~\cite{Giudice:2003jh,Buchmuller:2004nz} for the standard cosmology (SC)
case.

The structure of the paper is as follows. In Section II we briefly present the
basic formulae for the mechanism of thermal leptogenesis in braneworld
cosmology. In section III we present a simple fit for the efficiency factor as
a function of the decay parameter, obtained by solving numerically the
Boltzmann equations. In Section IV we find lower bounds on the decaying heavy
Majorana neutrino mass, the fundamental five-dimensional gravity scale and the
transition temperature from brane to standard cosmology. In Section V we derive
the upper bounds on the light neutrino masses implied by leptogenesis. Finally,
in section VI we present our conclusions.

\section{Thermal leptogenesis in the brane}

In its simplest standard model version, leptogenesis is dominated by the
$CP$-violating interactions of the lightest of the heavy Majorana neutrinos.
Assuming that right-handed neutrinos are hierarchical, $M_1 \ll M_{2,3}\,$, and
that the decays of the heavier neutrinos do not influence the final value of
the $B-L$ asymmetry, the baryon asymmetry generated by $N_1$ can be
conveniently parameterized as
\begin{align}
\eta_{B} = \frac{\xi}{f}~N_{B-L}^f~, \label{etab}
\end{align}
where $\xi=28/79$ is the fraction of the $B-L$ asymmetry converted into a
baryon asymmetry by sphaleron processes~\cite{Khlebnikov:sr} and $f=2387/86$ is
the dilution factor calculated assuming standard photon production from the
onset of leptogenesis till recombination.  The amount of $B-L$ asymmetry per
comoving volume, $N_{B-L}^f \equiv N_{B-L} (T \ll M_1)$, is obtained from the
solution of the relevant Boltzmann equations, which in the above minimal
framework can be written in the compact form~\cite{Buchmuller:2004nz},
\begin{align}
\label{BE1}
\frac{d N_{N_1}}{dz} =& - (D+S)\left(N_{N_1}-N_{N_1}^{eq}\right)~,\\
\label{BE2} \frac{d N_{B-L}}{dz} =& -\epsilon_1 D
\left(N_{N_1}-N_{N_1}^{eq}\right) - W N_{B-L}~,
\end{align}
where $z=M_1/T\,$, $N_{N_1}^{eq}$ is the equilibrium number density, normalized
so that $N_{N_1}^{eq}(z \ll 1) =3/4$, and $\epsilon_1$ is the $CP$-asymmetry
parameter in $N_1$ decays. Different physical processes contribute to these
equations. The terms $D$, $S$ and $W$, defined through the corresponding rates
$\mathit{\Gamma}_i$ as
\begin{align} \label{DSW}
D = \frac{\mathit{\Gamma}_D}{H z}\,,\quad S = \frac{\mathit{\Gamma}_S}{H
z}\,,\quad W = \frac{\mathit{\Gamma}_W}{H z}\,,
\end{align}
account for decays and inverse decays ($D$), $\Delta L = 1$ scatterings ($S$)
and washout processes ($W$), respectively. While decays yield the source for
the generation of the $B-L$ asymmetry, all other processes, including $\Delta L
= 2$ processes mediated by heavy neutrinos, contribute to the total washout
term.

During radiation domination $\rho=\pi^2 g_* T^4/30$ and the expansion rate
(\ref{Fried}) is given by
\begin{align}
H(z) =  \frac{\overline{H}}{z^2} \sqrt{\frac{1+ (z_t/z)^4}{1 + z_t^{4}}}~,
\label{Fried_rad}
\end{align}
where
\begin{align}
\overline{H} \equiv H(z=1) =\sqrt{\frac{4 \pi^3 g_*}{45}}
\frac{M_1^2}{M_P}\sqrt{1 + z_t^{4}}~,
\end{align}
and $g_*=106.75$ is the effective number of relativistic degrees of freedom.
The parameter $z_t$ is defined as
\begin{align} \label{zt}
z_t = \frac{M_1}{T_t} = \left( \frac{\pi^3 g_* M_P^2 M_1^4}{45
M_5^6}\right)^{1/4}~,
\end{align}
where $T_t$ is the temperature at which the transition from brane to standard
cosmology takes place ($\rho(T_t) \simeq 2\lambda$),
\begin{align}\label{Tt}
T_t \simeq 9.8 \times 10^7\,\text{GeV}
\left(\frac{M_5}{10^{12}\,\text{GeV}}\right)^{3/2}\,.
\end{align}

In establishing a connection between leptogenesis and neutrino physics, the
dependence of the different contributions given in Eq.~(\ref{DSW}) on the
neutrino parameters turns out to be crucial. In particular, one can show that
 terms $D$ and $S$ are proportional to the so-called effective neutrino mass
parameter
\begin{align}
\widetilde{m}_1 = \frac{(m_D^\dag m_D)_{11}}{M_1}\,,
\end{align}
where $m_D$ is the Dirac neutrino mass matrix. On the other hand, the washout
term $W$ contains in general two contributions, one which depends only on $\mt$
and another which is proportional to the absolute neutrino mass scale
$\overline{m}^2=m_1^2+m_2^2+m_3^2$.

Departure from thermal equilibrium is controlled by the decay parameter $K$,
\begin{align}\label{decaypar}
K = \frac{\widetilde{\mathit{\Gamma}}_D}{\overline{H}}
=\frac{\widetilde{m}_1}{m_*}\,,
\end{align}
where $\widetilde{\mathit{\Gamma}}_D = \mathit{\Gamma}_D(z=\infty)$ is the
$N_1$ decay width at zero temperature,
\begin{align}
\widetilde{\mathit{\Gamma}}_D = \frac{1}{8 \pi} \frac{\widetilde{m}_1
M_1^2}{v^2}\,,
\end{align}
$v \approx 174$~GeV and $m_*$ is the equilibrium neutrino mass
\begin{align}\label{mast}
 m_* = 8 \pi \overline{H} \frac{v^2}{M_1^2} \simeq 1.08 \times
 10^{-3}~\text{eV}\, \sqrt{1+z_t^4}\,.
\end{align}
In the high-energy regime of brane cosmology, i.e. for $z_t \gg 1$, we find
\begin{align}
m_* \simeq 1.1 \times 10^{-3}~\text{eV}\,\left(\frac{M_{1}}{10^{8}\,
\text{GeV}}\right)^{2} \left(\frac{10^{12}\,\text{GeV}}{M_{5}}\right)^{3}.
\end{align}

The solution of Eq.~(\ref{BE2}) for $N_{B-L}$ is given by
\begin{align}\label{NBL}
N_{B-L}(z) = N_{B-L}^i\,e^{-\int_{z_i}^z dz' W(z')}-\frac{3}{4}~ \epsilon_1~
\kappa(z)\,,
\end{align}
where $N_{B-L}^i$ accounts for any possible initial $B-L$ asymmetry (e.g. due
to the decays of the heavier neutrinos $N_{2,3}$) and $\kappa(z)$ is the
efficiency factor, which measures  $B-L$ production from $N_1$ decays,
\begin{align}
\kappa (z) = -\frac{4}{3} \int_{z_i}^z dz'\frac{D}{D+S}\frac{d
N_{N_1}}{dz'}~e^{-\int_{z'}^z dz'' W(z'')}\,.
\end{align}

Assuming no pre-existing asymmetry, i.e. $N_{B-L}^i=0$, from Eqs.~(\ref{etab})
and (\ref{NBL}) the final baryon asymmetry can be conveniently written as
\begin{align}
\eta_B = d\,\epsilon_1\,\kappa_f\,,
\end{align}
where $d \simeq 0.96 \times 10^{-2}$ and $\kappa_f=\kappa(\infty)$ is the final
value of the efficiency factor, normalized in such a way that $\kappa_f
\rightarrow 1$ in the limit of thermal initial $N_1$ abundance and no washout.
The computation of $\kappa_f$ is a rather difficult task, since it involves
solving  the Boltzmann equations. Nevertheless, simple analytical formulae can
be found in the limiting cases of strong ($K \gg 1$) and weak ($K \ll 1$)
washout regimes~\cite{Buchmuller:2004nz}, and numerical fits valid for all
values of $\mt$ can also be derived in the standard cosmology
case~\cite{Giudice:2003jh,Buchmuller:2004nz}. Clearly, one expects these
results to be modified in the braneworld scenario. In the next section we will
derive a simple fit for the efficiency factor, assuming that the baryon
asymmetry is created during the high-energy regime of brane cosmology.

\section{The efficiency factor}

Neutrino oscillation experiments presently constrain two mass squared
differences for the light neutrinos~\cite{Strumia:2005tc},
\begin{align}
m_{sol} &\equiv \sqrt{\Delta m_{sol}^2}\,\, \simeq 8.9 \times
10^{-3}\,\text{eV}\,,\nonumber\\
m_{atm} &\equiv \sqrt{\Delta m_{atm}^2} \simeq 5.0 \times 10^{-2}\,\text{eV}\,,
\label{nudata}
\end{align}
from solar and atmospheric measurements, respectively. The most stringent
bounds on the absolute neutrino mass scale come from $0\nu\beta\beta$-decay
experiments~\cite{Klapdor-Kleingrothaus:2000sn}, which yield an upper bound on
the light Majorana neutrino masses of about 1~eV, and from the 2df galaxy
redshift survey and WMAP results~\cite{Spergel:2003cb}, which indicate that
$\sum m_{i} < 0.7$~eV.

The effective neutrino mass parameter $\mt$ is an essential quantity for
thermal leptogenesis. Remarkably, it is possible to find a natural range for
this quantity in terms of the light neutrino masses. Indeed, the lower bound
$\mt \geq m_1$ is always verified~\cite{Fujii:2002jw}. Moreover, in the absence
of strong cancelations or specific fine-tunings, one has the upper bound $\mt
\lesssim m_3$~\cite{Buchmuller:2002rq}. Thus, one finds $m_1 \leq \mt \lesssim
m_3$.

\begin{figure}[t]
  \includegraphics[width=8.5cm]{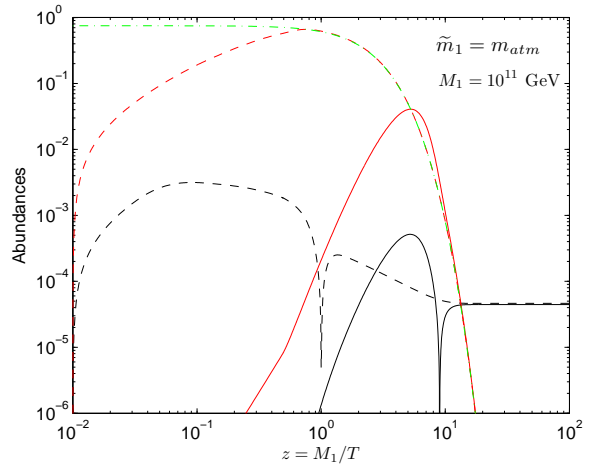}\\
  \caption{(Color online) The evolution of $N_{N_1}$ (red curves), $N_{N_1}^{eq}$
  (green curve) and $|\eta_B/\epsilon_1|$ (black curves) as functions of the
  temperature for $M_1=10^{11}$~GeV and  $\widetilde{m}_1 = m_{atm}$.
  The solid (dashed) lines correspond to the brane (standard) cosmology result.}
  \label{fig1}
\end{figure}

Of particular interest is the favored neutrino mass range $m_{sol} \lesssim \mt
\lesssim m_{atm}$. In this situation, leptogenesis in the standard cosmology
case would occur in the mildly strong washout regime $8 \lesssim K \lesssim 46$
(see Eqs.~(\ref{decaypar}) and (\ref{mast})). On the other hand, in the high
energy regime of brane cosmology ($z_t \gg 1$) one has $8/z_t^2 \lesssim K
\lesssim 46/z_t^2\,$, which for $z_t > 7$ already implies that leptogenesis in
the brane lies in the weak washout regime. As an illustrative example, in
Fig.~\ref{fig1} we compare the evolution of the abundances with temperature in
the SC strong  and BC weak washout regimes. Assuming zero initial $N_1$
population at $T \gg M_1$, we have numerically solved the system of Boltzmann
equations (\ref{BE1}) and (\ref{BE2}) following the approach presented in
Ref.~\cite{Giudice:2003jh}, which properly includes the leading
finite-temperature corrections and takes into account all the relevant
processes. The evolution of the $N_1$ abundance (red curves) and the ratio
$|\eta_B/\epsilon_1|$ (black curves) are shown as functions of $z$ for
$M_1=10^{11}$~GeV and $\widetilde{m}_1 = m_{atm} \simeq 0.05$~eV. The dashed
lines correspond to the standard cosmology result, while the solid lines are
the brane cosmology result, taking $M_5 = 4.5\times 10^{12}$~GeV. With this
choice of parameters, $K_{\rm SC} = 46$ and $K_{\rm BC}=4 \times 10^{-3}$.

\begin{figure} [t]
\includegraphics[width=8.5cm]{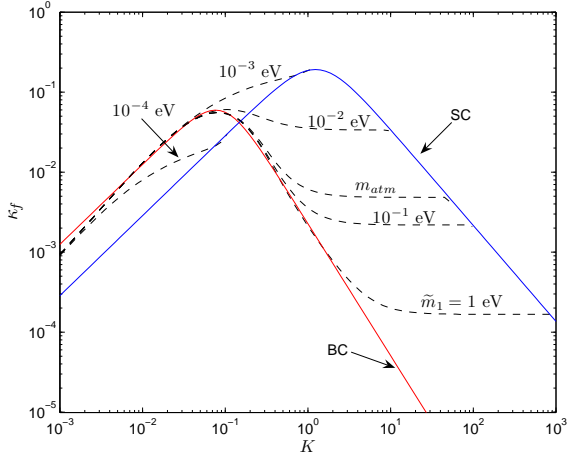}\\
\caption{(Color online) The efficiency factor $\kappa_f$ as a function of the
decay parameter $K$, obtained by solving numerically the Boltzmann equations,
for different values of the effective neutrino mass $\mt$. The blue and red
solid curves correspond to the numerical fits of $\kappa_f$ given in
Eqs.~(\ref{kfinterp}),
 (\ref{fitSC}) and (\ref{fitBC}), for the cases of standard and brane cosmology,
respectively.}\label{fig2}
\end{figure}

It is possible to find a simple numerical fit for the efficiency factor
$\kappa_f$ in terms of the decay parameter $K$. In both SC and BC cases, the
behavior of $\kappa_f$ is well described by the power law fit
\begin{align}
\kappa_f^{-1} = \frac{a}{K} + \left(\frac{K}{b}\right)^{c},\label{kfinterp}
\end{align}
valid for a heavy neutrino mass $M_1 \lesssim 10^{14}$~GeV. For SC we find
\begin{align}
a=3.5\,,\,b=0.6\,,\,c=1.2\,, \label{fitSC}
\end{align}
(see also Ref.~\cite{Giudice:2003jh}), while in the high-energy regime of BC we
obtain
\begin{align}
a=0.8\,,\,b=0.025\,,\,c=1.65\,. \label{fitBC}
\end{align}
These fits are plotted in Fig.~\ref{fig2} and compared with the efficiency
factor obtained by solving numerically the full set of Boltzmann equations for
different values of ${\widetilde{m}_1}$. As expected, the curves in the BC case
are displaced to the weak washout region where thermal leptogenesis is more
efficiently realized.

Using the numerical fit (\ref{kfinterp}), one can see that maximal efficiency
is achieved at
\begin{align}
K_{\rm peak} = \left(\dfrac{a\,b^{\,c}}{c}\right)^{1/(c+1)}~,
\end{align}
which in BC corresponds to
\begin{align} \label{kfpeakBC}
K_{\rm peak} \simeq 0.08\,,\quad \kappa_{f}^{\rm peak} \simeq 0.06\,,
 \end{align}
while in SC one finds
\begin{align} \label{kfpeakSC}
K_{\rm peak} \simeq 1.23\,, \quad \kappa_{f}^{\rm peak}\simeq 0.19\,.
\end{align}
We notice that this value of $\kappa_f$ in the standard cosmology case implies
an effective neutrino mass $\mt \simeq 1.3 \times 10^{-3}$~eV. On the other
hand, the maximal $\kappa_f^{\rm peak}$ in the high-energy regime of BC can be
achieved for any value of $\mt \gtrsim 10^{-3}$~eV, including the favored
neutrino mass range $m_{sol} \lesssim \mt \lesssim m_{atm}$.

The above values for the maximal efficiency factor together with the upper
bound on the $CP$ asymmetry $\epsilon_1$ can be translated into bounds on the
heavy Majorana neutrino mass $M_1\,$, the 5D Planck scale $M_5\,$, the
transition temperature $T_t$ and the light neutrino masses $m_i\,$.

\section{Lower bounds on $M_1\,$, $M_5$ and $T_t$}

In order to discuss the bounds implied by leptogenesis, it is convenient to
write the $CP$ asymmetry $\epsilon_1$ as the product
\begin{align}
\epsilon_1 =\epsilon_1^{\rm max} \sin \delta_L\,,
\end{align}
where $\epsilon_1^{\rm max}$ is the maximal $CP$ asymmetry and $\delta_L$ is an
effective leptogenesis phase. In general, $\epsilon_1^{\rm max}$ depends on
$M_1$, $\mt$ and the light neutrino masses~\cite{Buchmuller:2002rq}. In the
case of hierarchical and quasi-degenerate light neutrinos,
\begin{align} \label{eps1max}
\epsilon_1^{\rm max} = \frac{3}{16\pi}\frac{M_1\,m_{3}}{v^2}
\left[1-\frac{m_1}{m_3}\sqrt{1+\dfrac{m_3^2-m_1^2}{\mt^2}}\right]\,,
\end{align}
where for a normal hierarchy
\begin{align}
m_3^2 &=m_1^2+\,m_{atm}^2+\,m_{sol}^2\,,\nonumber\\
m_2^2 &=m_1^2+m_{sol}^2\,,\\
\overline{m}^2 &=3\,m_1^2+m_{atm}^2+2\,m_{sol}^2\,.\nonumber
\end{align}
The $CP$-asymmetry reaches its maximum for $m_1=0$ and $m_3 \simeq m_{atm}\,$,
i.e. for fully hierarchical neutrinos. In this case,
\begin{eqnarray}
\epsilon_1^{\rm max}&=&\frac{3}{16\pi}\frac{M_1\,m_{atm}}{v^2}\nonumber\\
&\simeq & 10^{-6}\,\left(\frac{M_1}{10^{10}\,{\rm GeV}} \right)
\,\left(\frac{m_{atm}}{0.05\,{\rm eV}}\right)\,.
\end{eqnarray}

A lower bound on $M_1$ can be obtained from the condition
\begin{align}
\label{condition} \eta_{B}^{\rm max}= d\,\epsilon_1^{\rm max}\,\kappa_{f}^{\rm
peak} \geq \eta_{B}^{\rm obs},
\end{align}
where $\eta_{B}^{\rm obs}$ is given in Eq.~(\ref{BAU}) and $\eta_{B}^{\rm max}$
is the maximal baryon asymmetry with the efficiency factor $\kappa_{f}^{\rm
peak}$ given in Eqs.~(\ref{kfpeakBC}) and (\ref{kfpeakSC}) for the BC and SC
regimes, respectively.

The condition (\ref{condition}) then leads to the following lower bound on
$M_1$
\begin{align}
\label{boundM1a} M_1 \gtrsim 6.3\times 10^{8}\,{\rm GeV}
\left(\frac{\eta_B^{\text{obs}}}{6\times 10^{-10}}\right)
\left(\frac{0.05\,{\rm eV}}{m_{atm}}\right)~(\kappa_{f}^{\rm peak})^{-1}.
\end{align}

Taking into account the minimum value for $\eta_B$ allowed by observations
$\eta_B^{\rm obs}=5.8 \times 10^{-10}$ (cf. Eq.~(\ref{BAU})) and the
atmospheric neutrino mass difference result given in Eq.~(\ref{nudata}), one
obtains
\begin{align}
\label{boundM1b} M_1 \gtrsim 6.1 \times 10^8\,{\rm GeV}\,(\kappa_{f}^{\rm
peak})^{-1} ,
\end{align}
In the SC leptogenesis regime, $\kappa_{f}^{\rm peak} \simeq 0.19$ and one gets
$M_1 \gtrsim 3.2 \times 10^9\,{\rm GeV}$. On the other hand, in the high-energy
regime of brane cosmology, $\kappa_{f}^{\rm peak} \simeq 0.06$ and one obtains
the more restrictive bound
\begin{align}
\label{boundM1f} M_1 \gtrsim 1.0 \times 10^{10}~~{\rm GeV}.
\end{align}

This bound implies, in turn, a lower bound on the gravity scale $M_5$. Indeed,
imposing $K=K_{\rm peak}$, with $K$ given in Eq.~(\ref{decaypar}), and using
Eq.~(\ref{zt}) and the bound~(\ref{boundM1f}), we get
\begin{align}
\label{boundM5} M_5 \gtrsim 9.7\times 10^{12}\,{\rm GeV}
 \left(\frac{\mt}{10^{-3}\,{\rm eV}}\right)^{-1/3}.
\end{align}
Finally, from this result and using Eq.~(\ref{Tt}), we can derive the following
lower bound on the transition temperature
\begin{align} \label{boundTt}
T_t \gtrsim 3.0 \times 10^9\,{\rm GeV}
 \left(\frac{\mt}{10^{-3}\,{\rm eV}}\right)^{-1/2}.
\end{align}

In the favored neutrino mass range $m_{sol} \lesssim \mt \lesssim m_{atm}$, the
bounds (\ref{boundM5}) and (\ref{boundTt}) imply
\begin{align}
M_5 & \gtrsim (2.6 - 4.7) \times 10^{12}~{\rm GeV}\,,\\
T_t & \gtrsim (4.2 - 9.9) \times 10^{8}~{\rm GeV}\,.
\end{align}
For $\mt \lesssim 1$~eV, these bounds yield $M_5 \gtrsim 1.0 \times
10^{12}\,{\rm GeV}$ and $T_t \gtrsim 1.0 \times 10^8\,{\rm GeV}$.

\section{Upper bound on the light neutrino masses}

The washout term $W$ receives contributions from inverse decays, $\Delta L =1$
processes and $\Delta L = 2$ processes. At low temperatures ($z \gg 1$) the
contribution $\Delta W$, which depends on the absolute neutrino mass scale
$\overline{m}\,$, dominates over the contribution proportional to
$\widetilde{m}_1$. This contribution is approximately given by
\begin{align}
\Delta W(z \gg 1) \simeq \frac{c_W}{z^2 \sqrt{1+
    (z_t/z)^4}}\,,\label{deltawashout}
\end{align}
where~\cite{Buchmuller:2004nz}
\begin{align}
c_W = \frac{9 \sqrt{5}\, M_P\,M_1\,\overline{m}^2}{8\, \pi^{9/2}
\sqrt{g_*}\,v^4} \simeq 0.19 \left(\frac{M_1}{10^{10}\,\rm{GeV}}\right)
\left(\frac{\overline{m}}{\rm{1\,eV}}\right)^2\,.
\end{align}

The presence of $\Delta W$ modifies the total efficiency factor,
\begin{align}
\overline{\kappa}_f = \kappa_f\, e^{-\int^\infty_{z_B}dz\,\Delta W}\,,
\end{align}
where $z_B$ is the value at which the lepton asymmetry is no longer produced.
The washout term in Eq.~(\ref{deltawashout}) leads then to
\begin{align}
\overline{\kappa}_f  = \kappa_f\, e^{-c_W \, f(z_B,z_t)}\,,
\end{align}
with
\begin{align}
 f(z_B,z_t) &= \int^\infty_{z_B}dz\,\dfrac{1}{z^2 \sqrt{1+
    (z_t/z)^4}}\\
&=\frac{1}{4\,\sqrt{i}\, z_t}\,
B\left[-\left(\frac{z_t}{z_B}\right)^4;\frac{1}{4},\frac{1}{2}\right]\,,
\end{align}
where $B(z;a,b)$ is the incomplete beta function. For $z_t \ll z_B$ one
recovers the standard cosmology result~\cite{Buchmuller:2004nz}
\begin{align}
f(z_B) \simeq \frac{1}{z_B}\,,
\end{align}
while for $z_t \gg z_B$ (BC regime) one finds
\begin{align}
f(z_t) \simeq \frac{\Gamma(1/4)\,\Gamma(5/4)}{\sqrt{\pi}\,z_t} \simeq
\dfrac{1.85}{z_t}\,,
\end{align}
leading to
\begin{align} \label{kappafbar}
\overline{\kappa}_f \simeq \kappa_f\, \exp\,\left[-3.4 \times 10^{-3}
\left(\frac{M_5}{10^{12}\,\rm{GeV}}\right)^{3/2}
\left(\frac{\overline{m}}{1\,\rm{eV}}\right)^2\right]\,.
\end{align}

\begin{figure}[t]
  \includegraphics[width=9cm]{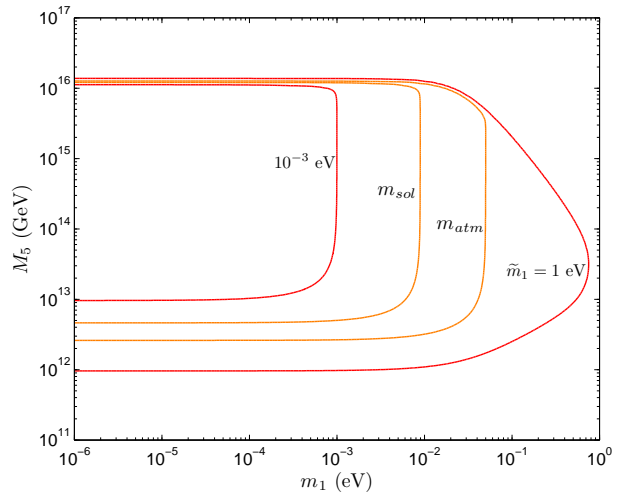}
  \caption{(Color online) Contour lines of constant
  $\eta_B^{\rm max}=\eta_B^{\rm obs}=5.8 \times 10^{-10}$ in the
  $(m_1\,,M_5)$-plane for different values of $\mt\,$. We assume that
  thermal leptogenesis occurs in the high-energy regime of brane cosmology.}\label{fig3}
\end{figure}

We notice that the dependence of the washout term on $M_1$ is canceled out in
the BC regime. In the latter case, the $\Delta W$ contribution to the total
washout depends only on the fundamental scale $M_5$ and the absolute neutrino
mass scale $\overline{m}$. Clearly, the exponential suppression in the
efficiency factor becomes relevant only for large values of $M_5$ or
$\overline{m}$. In Fig.~\ref{fig3} we present the contour lines of constant
$\eta_B^{\rm max}=\eta_B^{\rm obs}=5.8 \times 10^{-10}$ in the
$(m_1\,,M_5)$-plane for different values of $\mt\,$. The baryon asymmetry
$\eta_B^{\rm max}$ was computed using the maximal $CP$-asymmetry (cf.
Eq.~(\ref{eps1max})) and taking into account the efficiency factor
$\overline{\kappa}_f$ as given in Eq.~(\ref{kappafbar}) with $\kappa_f$ at its
peak value (\ref{kfpeakBC}). From the figure one finds the upper bound $M_5
\lesssim 10^{16}$~GeV, which translates into an upper bound on the transition
temperature, $T_t \lesssim 10^{14}$~GeV. On the other hand, the light neutrino
mass $m_1$ turns out to be constrained essentially  by the value of $\mt$. For
$\mt \leq 1$~eV, we find $m_1 < 0.8$~eV. Thus, one concludes that the thermal
leptogenesis scenario in braneworld cosmology is consistent with
quasi-degenerate light neutrinos.

\section{Conclusions}

The explanation of the matter-antimatter asymmetry observed in the universe
remains an open question. Leptogenesis is a simple and elegant mechanism for
explaining the cosmological baryon asymmetry. Since this asymmetry must has
been created at early times, a successful realization of this mechanism
crucially depends on the properties of early cosmology.

In this paper, we have studied the implications of Randall-Sundrum type II
braneworld cosmology on the thermal leptogenesis scenario. In contrast with the
usual leptogenesis scenario of standard cosmology, where low-energy neutrino
data favors a strong washout regime and quasi-degenerate light neutrinos are
not compatible with leptogenesis, we have found that in the high-energy regime
of braneworld cosmology leptogenesis can be successfully realized in a weak
washout regime, and a quasi-degenerate neutrino mass spectrum is allowed.

For an initially vanishing heavy Majorana neutrino abundance, we have obtained
bounds on the decaying heavy Majorana neutrino mass, the fundamental
five-dimensional gravity scale and the transition temperature from brane to
standard cosmology. As far as the upper bound on the light neutrino masses is
concerned, we have seen that thermal leptogenesis in the brane imposes a limit
less restrictive than $m_i < 0.23$~eV, which is the one presently implied by
WMAP results. Finally, it is worth emphasizing that all the bounds have been
derived under the simple assumption of hierarchical heavy Majorana neutrinos.
For a partially degenerate spectrum these bounds can be relaxed. Indeed, if at
least two of the heavy Majorana neutrinos are quasi-degenerate in mass, i.e.
$M_1 \approx M_2$, then the leptonic $CP$ asymmetry relevant for leptogenesis
exhibits the resonant behavior $\epsilon_1 \sim
M_1/(M_2-M_1)$~\cite{Flanz:1994yx}, and values of $\epsilon_1^{\rm max} \sim
\mathcal{O}(1)$ can be reached. In this case, the lower bound on $M_1$ coming
from leptogenesis can be relaxed~\cite{Giudice:2003jh,Buchmuller:2004nz}.

Let us also briefly comment on the consistency of inflation and reheating with
the present framework of thermal leptogenesis. Assuming that inflation is
driven by a simple quadratic chaotic inflation potential, $V=\frac{1}{2} m^2
\varphi^2$, the scale of inflation in the $\rho^2$-dominated period is roughly
given by $V \sim (0.1\,M_5)^4$~\cite{Maartens:1999hf}, thus ensuring that
sufficient inflation can take place while fulfilling the requirement
$V<M_5^4\,$. Another important constraint comes from the reheating process,
often associated with the decay of the inflaton into elementary particles, and
which eventually results in the creation of a thermal bath. If the inflaton
decays through Yukawa interactions with a coupling to matter fields of order
$\sim \mathcal{O}(1)$, then the decay rate is fast, $\Gamma_\varphi \simeq m$,
and reheating is practically instantaneous. Assuming that inflation ends in the
$\rho^2$-regime, the reheating temperature is estimated as $T_{rh} \sim
\mathcal{O}(10^{-2}) \times M_5\,$, so that the energy density of radiation
$\rho_{rh}=\pi^2 g_* T_{rh}^4/30 < M_5^4\,$. Since consistency with braneworld
thermal leptogenesis requires $T_t \lesssim M_1 \lesssim T_{rh}\,$, from
Eq.~(\ref{Tt}) one obtains the upper bound $M_5 \lesssim 10^{16}$~GeV, which
coincides with the leptogenesis bound implied by the observed baryon asymmetry
(cf. Fig.~3). Notice however that, for values of $M_5$ close to the upper
limit, i.e. $M_5 \sim 10^{16}$~GeV, one would have $T_t \simeq M_1 \simeq
T_{rh} \simeq 10^{14}$~GeV and the mechanism of leptogenesis would occur during
the transition from brane to standard cosmology.

During the reheating process the distribution of particles is far from thermal
equilibrium. Thus, number-conserving as well as number-violating interactions
are necessary to achieve kinetic and chemical equilibrium among the different
species. In an expanding universe, full thermal equilibrium requires a
sufficiently low temperature. Indeed, $2 \leftrightarrow 2$ GUT interactions
mediated by massless gauge bosons would have occurred in the very early
universe at a rate $\Gamma \sim \alpha^2_{\rm GUT}\,T$, with $\alpha_{\rm GUT}
\sim \mathcal{O}(10^{-2})$. Since during the radiation-dominated era of
standard cosmology $H \sim T^2/M_P\,$, such interactions are expected to be in
equilibrium for $T \lesssim \alpha^2_{\rm GUT}\,M_P \sim
10^{15}$~GeV~\cite{Ellis:1979nq}. On the other hand, in the high-energy regime
of braneworld cosmology the expansion rate scales as $H \sim T^4/M_5^3$ and,
consequently, thermal equilibrium is expected at $T \lesssim \alpha^{2/3}_{\rm
GUT}\, M_5 \lesssim 5\times 10^{14}$~GeV for $M_5 \lesssim 10^{16}$~GeV.
Clearly, the above analysis could be subject to modifications, if there exist
new interactions which could thermalize the universe at the earliest epochs.
For instance, it is conceivable that quantum gravity effects near the
fundamental Planck scale could have resulted in a state of maximal entropy.
Nevertheless, in the absence of a complete theory of quantum gravity, this
remains an open issue.

\bigskip

\emph{Note added:} While this work was in preparation, a related preprint has
appeared~\cite{Okada:2005kv}. However, the conclusions drawn by the authors
about mass scales and parameters somewhat differ from ours because in
Ref.~\cite{Okada:2005kv} the corresponding bounds were derived by using the
right-handed neutrino thermalization condition instead of solving the full set
of Boltzmann equations.

\begin{acknowledgments}
We thank F.~R.~Joaquim for reading the manuscript. The work of R.G.F and
N.M.C.S. was supported by Funda\c{c}\~{a}o para a Ci\^{e}ncia e a Tecnologia (FCT,
Portugal) under the grants SFRH/BPD/1549/2000 and SFRH/BD/4797/2001,
respectively. This work has been partially supported by FCT through the
projects POCI/FIS/56093/2004, POCTI/FNU/44409/2002 and PDCT/FP/FNU/50250/2003.
\end{acknowledgments}

\end{document}